\pdfoutput=1

\documentclass{PoS1}
\usepackage{amsmath,amssymb,amsbsy,cite}
\usepackage{graphicx}
\usepackage{epsfig}
\usepackage{rotating}

\def\tabvspace{\vphantom{$\Big($}}

\newcommand{\Delstar}{\ensuremath{\Delta^{\raise0.18ex\hbox{${\scriptstyle *}$}}}}
\def\gtwid{{\,\raise.35ex\hbox{$>$\kern-.75em\lower1ex\hbox{$\sim$}}\,}}
\def\ltwid{{\,\raise.35ex\hbox{$<$\kern-.75em\lower1ex\hbox{$\sim$}}\,}}
\def\leftvec{{\raise1.5ex\hbox{$\leftarrow$}\kern-1.00em}}
\def\rightvec{{\raise1.5ex\hbox{$\rightarrow$}\kern-1.00em}}
\def\half{{\scriptstyle \raise.2ex\hbox{${1\over2}$}}}
\def\threehalves{{\scriptstyle \raise.15ex\hbox{${3\over2}$}}}
\def\third{{\scriptstyle \raise.15ex\hbox{${1\over3}$}}}
\def\third{{\scriptstyle \raise.15ex\hbox{${1\over3}$}}}
\def\twothirds{{\scriptstyle \raise.15ex\hbox{${2\over3}$}}}
\def\fourth{{\scriptstyle \raise.15ex\hbox{${1\over4}$}}}

\newcommand*{\bea}{\begin{eqnarray}}
\newcommand*{\eea}{\end{eqnarray}}
\newcommand*{\be}{\begin{equation}}
\newcommand*{\ee}{\end{equation}}

\def\mev{{\rm MeV}}

\def\Wslash{W\hskip-0.65em /}

\newcommand*{\CPT}{\raise0.4ex\hbox{$\chi$}PT}
\newcommand*{\chpt}{\raise0.4ex\hbox{$\chi$}PT}
\newcommand*{\schpt}{S\raise0.4ex\hbox{$\chi$}PT}

\def\Standard Modelb{\sigma_M}
\def\Standard Modelt{\overline{\sigma}_M}

\def\eqref#1{{(\ref{#1})}}

\def\bar{\overline}
\def\hat{\widehat}
\def\tilde{\widetilde}

\def\bea{\begin{eqnarray}}
\def\eea{\end{eqnarray}}

\def\beq{\begin{equation}}
\def\eeq{\end{equation}}

\def\spose#1{\hbox to 0pt{#1\hss}}
\def\ltapprox{\mathrel{\spose{\lower 3pt\hbox{$\mathchar"218$}}
 \raise 2.0pt\hbox{$\mathchar"13C$}}}
\def\gtapprox{\mathrel{\spose{\lower 3pt\hbox{$\mathchar"218$}}
 \raise 2.0pt\hbox{$\mathchar"13E$}}}
\def\inapprox{\mathrel{\spose{\lower 3pt\hbox{$\mathchar"218$}}
 \raise 2.0pt\hbox{$\mathchar"232$}}}

\title{Flavor Physics in the LHC era: the role of the lattice}

\ShortTitle{Flavor Physics in the LHC era: the role of the lattice}

\author{Jack Laiho\\
        Department of Physics and Astronomy, University of Glasgow, Glasgow, G128 QQ, UK\\
        E-mail: \email{jlaiho@fnal.gov}}

\author{\speaker{Enrico Lunghi}\\
       Indiana University, Bloomington, IN 47405 \\
       E-mail: \email{elunghi@indiana.edu}}

\author{Ruth S. Van de Water\\
        Physics Department, Brookhaven National Laboratory, Upton, NY 11973\\
        E-mail: \email{ruthv@bnl.gov}}

\abstract{We discuss the present status of global fits to the CKM unitary triangle using the latest experimental and theoretical constraints.  For the required nonperturbative weak matrix elements, we use three-flavor lattice QCD averages from \texttt {www.latticeaverages.org}; these have been updated from Ref.~\cite{Laiho:2009eu} to reflect all available lattice calculations as of the ``{\it End of 2011}.''  Because of the greater than $3\sigma$ disagreement between the extraction of $|V_{ub}|$ from inclusive and exclusive semileptonic $b\to u \ell\nu \;(\ell = e,\mu)$ decays, particular emphasis is given to a clean fit in which we remove the information from these decays.   Given current theoretical and experimental inputs, we observe an approximately 3$\sigma$ tension in the CKM unitarity triangle that may indicate the presence of new physics in the quark-flavor sector.  Using a model-independent parameterization of new-physics effects, we test the compatibility of the data with scenarios in which the new physics is in kaon mixing, in $B$-mixing, or in $B\to\tau\nu$ decay.  We find that scenarios with new physics in $B\to\tau\nu$ decay or $B$-mixing are approximately equally preferred.  If we interpret these results in terms of contributions to $\Delta S = 2$ and $\Delta B = 2$ four-fermion operators, we find that the preferred scale of new physics (with Standard-Model like couplings) is in the few hundred GeV range.}

\FullConference{XXIX International Symposium on Lattice Field Theory \\
		 July 10--16 2011\\
		 Squaw Valley, Lake Tahoe, California}

\begin{document}
\section{Motivation}
The $B$-factories and the Tevatron have produced a remarkable wealth of data needed to determine elements of the Cabibbo-Kobayashi-Maskawa (CKM) matrix and to search for new physics beyond the Standard Model CKM framework.  Despite the great experimental success of the Standard Model, however, there is now considerable evidence for physics beyond the Standard Model, such as dark matter, dark energy, and neutrino masses.  Generic new-physics models to explain such phenomena also lead to additional $CP$-violating phases beyond the single one in the Standard Model;  this would lead to apparent inconsistencies between independent determinations of the CKM matrix elements.  Although there is presently reasonably good agreement with the Standard Model prediction of a single $CP$-violating phase, as measured by global fits of the CKM unitarity triangle, some tensions have been observed~\cite{Lunghi:2007ak, Lunghi:2008aa, Buras:2008nn, Buras:2009pj, Lenz:2010gu,Bona:2009cj,Lunghi:2009sm}. In this work we use the latest theoretical and experimental inputs to test the Standard-Model CKM framework in the quark flavor sector.  We first use a global fit to the CKM unitarity triangle to quantify the tension with the Standard Model, and then identify within a largely model-independent framework the most likely sources of the new physics.

\section{Unitarity Triangle Fit Preliminaries}
\label{sec:Prelim}

\subsection{Inputs}
\label{sec:inputs}
The standard analysis of the unitarity triangle involves a simultaneous fit to several quantities:  $\varepsilon_K$, $\Delta M_{B_d}$, $\Delta M_{B_s}$, time--dependent $CP$ asymmetry in $B\to J/\psi K_s$ ($S_{\psi K} = \sin (2\beta)$, where $\beta$ is the phase of $V_{td}^*$),\footnote{A discussion of penguin pollution in $S_{\psi K}$ can be found in Ref.~\cite{Lunghi:2010gv}; see also Refs.~\cite{Boos:2004xp, Li:2006vq, Bander:1979px, Gronau:2008cc, Ciuchini:2005mg, Faller:2008zc, Ciuchini:2011kd}.} direct $CP$ asymmetries in $B\to D^{(*)} K^{(*)}$ ($\gamma$ is the phase of $V_{ub}^*$) time dependent $CP$ asymmetries in $B\to (\pi\pi, \rho\rho, \rho\pi)$ ($\alpha = \pi-\beta-\gamma$), ${\rm BR} (B\to\tau\nu)$,  $|V_{ub}|$ and $|V_{cb}|$ (from both inclusive and exclusive $b\to (u,c)\ell\nu$ with $\ell = e,\mu$). We summarize the relevant inputs required for this analysis in Table~\ref{inputs}.

There are several choices for how to implement the constraints from $B\to\tau\nu$ leptonic decay and $B_{d,s}$ mixing ($\Delta M_{B_d}$ and $\Delta M_{B_s}$) on the unitarity triangle because one can parameterize the nonperturbative weak matrix element contributions to these quantities in different ways.  Certain combinations of lattice inputs are preferable, however, because they minimize correlations between the three different unitarity triangle constraints so that they can safely be neglected in the global fit. Let us now summarize the main considerations that lead to a reasonable choice of uncorrelated inputs. First of all it is important to include only one input with mass dimension 1 in order to eliminate correlations due to the determination of the lattice scale. Another important consideration is that the largest source of uncertainty in the $SU(3)$--breaking ratios, $\xi$ and $f_{B_s}/f_{B_d}$, is the chiral extrapolation.  Because the chiral logarithms are larger when the quark masses are lighter, the chiral extrapolation in the $SU(3)$-breaking ratios is primarily due to the chiral extrapolation in the $B_d$ quantities ({\it i.e.} $f_{B_d}$ and $\hat B_d$).  These ratios are, therefore, more correlated with $B_d$ rather than with $B_s$ quantities. Finally we note that the decay constants $f_{B_d}$ and $f_{B_s}$ have a stronger chiral extrapolation than the $B$-parameters $\hat B_d$ and $\hat B_s$.  In view of these considerations we choose to describe $B_q$ mixing in terms of $f_{B_s}\sqrt{\hat B_s}$ and $\xi$.  We then need one additional input to describe ${\rm BR} (B\to\tau\nu)$.  Although in principle the choice of $\hat B_d$ is preferable because $f_{B_d}$ has mass dimension and is more correlated with $\xi$ through statistics and the chiral extrapolation, we choose to use $f_{B_d}$ anyway.  This is because two (mostly) independent 2+1 flavor determinations of $f_{B_d}$ are currently available while only one group has presented the calculation of the bag parameters $\hat B_d$. Further, $f_{B_d}$ is known much more precisely and hence places a stronger constraint on the unitarity triangle.

In our analysis, we write 
\begin{equation}
\Delta M_{B_d} \propto \left(\frac{ f_{B_s} \sqrt{\hat B_s}}{\xi}\right)^2 \; , \quad \quad
\Delta M_{B_s} \propto \left(f_{B_s} \textstyle \sqrt{\hat B_s}\right)^2 \; ,\quad \quad
{\rm BR} (B\to \tau \nu) \propto f_{B_d}^2 \; .
\end{equation}
For completeness we point out that there is an alternative choice of inputs ($f_{B_s}/f_{B_d}$, $\hat B_s/\hat B_d$, $f_{B_s}$ and $\hat B_s$) for which correlations are again fairly small.
\begin{table}[t]
\begin{center}
\setlength{\unitlength}{1cm}
\begin{picture}(1, 1)
\put(-7.5, -2.6){ \begin{sideways} Lattice inputs \end{sideways}}
\put(-7.5, -6.8){ \begin{sideways} Other inputs \end{sideways}}
\end{picture}
\begin{tabular}{ll}
\hline
\hline
$\left| V_{cb} \right|_{\rm excl} =(39.5 \pm 1.0) \times 10^{-3}$~\cite{Okamoto:2004xg,Bailey:2010gb,Asner:2010qj}&
$\left| V_{ub} \right|_{\rm excl} = (3.12 \pm 0.26) \times 10^{-3} $~\cite{Dalgic:2006dt,Bailey:2008wp,delAmoSanchez:2010af,Ha:2010rf} \tabvspace \\
$\hat B_K = 0.7674 \pm 0.0099$~\cite{Durr:2011ap,Gamiz:2006sq,Laiho:2011dy,Kelly:2012uy,Bae:2011ff} &
$\kappa_\varepsilon = 0.94 \pm 0.01$~\cite{Goode:2011pd} \tabvspace \\
$\xi  \equiv f_{B_s}\sqrt{\hat B_s}/(f_{B_d}\sqrt{\hat B_d}) = 1.237 \pm 0.032$~\cite{Evans:2008zzg,Gamiz:2009ku,Albertus:2010nm} &
$f_{B_d} = (190.6 \pm 4.6) \; \mev$~\cite{Bazavov:2011aa,Na:2012kp} \tabvspace \\
$f_{B_d} \sqrt{\hat B_{B_d}} = (227 \pm 17) \; \mev $~\cite{Bouchard:2011xj,Gamiz:2009ku} & 
$f_{B_s} \sqrt{\hat B_{B_s}} = (279 \pm 13) \; \mev $~\cite{Bouchard:2011xj,Gamiz:2009ku} \tabvspace \\
$f_K = (156.1 \pm 1.1) \; \mev $~\cite{Davies:2010ip,Laiho:2011dy,Bazavov:2010hj,Kelly:2012uy} & \tabvspace\\
\hline\hline
$\left| V_{cb} \right|_{\rm incl} =(41.68 \pm 0.44 \pm 0.09 \pm 0.58) \times 10^{-3}$~\cite{Asner:2010qj}&
$\alpha = (89.5 \pm 4.3)^{\rm o}$ \tabvspace\\
$\left| V_{ub} \right|_{\rm incl} = (4.34 \pm 0.16^{+0.15}_ {-0.22}) \times 10^{-3} $~\cite{Asner:2010qj}  &
 $\eta_1 = 1.87 \pm 0.76$~\cite{Herrlich:1993yv,Brod:2011ty}  \vphantom{\Big(} \\
${\rm BR} (B\to \tau\nu) = (1.68 \pm 0.31) \times 10^{-4}$~\cite{Ikado:2006un,Sanchez:2010rt,Hara:2010dk} & 
$S_{\psi K_S} = 0.668 \pm 0.023$~\cite{Kreps:2010ts} \\
$\Delta m_{B_d} = (0.508 \pm 0.004)\; {\rm ps}^{-1}$~\cite{Asner:2010qj} & 
$\gamma = (78 \pm 12)^{\rm o}$~\cite{Bona:2005vz,Bona:2006ah} \vphantom{\Big(} \\
$\Delta m_{B_s} = (17.78 \pm 0.12 )\;  {\rm ps}^{-1}$~\cite{Asner:2010qj} & 
$\eta_2 = 0.5765 \pm 0.0065$~\cite{Buras:1990fn}  \vphantom{\Big(}\\
$m_{t, pole} = (173.2 \pm 0.9) \; {\rm GeV}$~\cite{Potamianos:2012sk} & 
$\eta_3 = 0.496 \pm 0.047$~\cite{Herrlich:1995hh,Brod:2010mj}  \vphantom{\Big(}\\
$m_c(m_c) = (1.273 \pm 0.006 ) \; {\rm GeV}$~\cite{McNeile:2010ji}&  
$\eta_B = 0.551 \pm 0.007$~\cite{Buchalla:1996ys} \vphantom{\Big(} \\
$\varepsilon_K = (2.229 \pm 0.012 ) \times 10^{-3}$~\cite{Yao:2006px} &
$\lambda = 0.2255  \pm 0.0007$~\cite{Antonelli:2008jg}\vphantom{\Big(}  \\ 
\hline
\hline
$\left| V_{cb} \right|_{\rm avg} =(40.77 \pm 0.81) \times 10^{-3}$&
$\left| V_{ub} \right|_{\rm avg} = (3.74 \pm 0.59) \times 10^{-3} $ \tabvspace \\
\hline
\hline
\end{tabular}
\caption{Lattice-QCD and other inputs to the unitarity triangle analysis. The determination of $\alpha$ is obtained from a combined isospin analysis of $B\to (\pi\pi,\; \rho\rho, \; \rho\pi)$ branching ratios and $CP$ asymmetries~\cite{Asner:2010qj}. Details on the lattice-QCD averages are given in the "Note on the correlations between the various lattice calculations" at $\texttt{www.latticeaverages.org}$.  Recent summaries of lattice-QCD progress on pion, kaon, charm, and bottom physics can be found in Refs.~\cite{Wittig:2012ha,Mawhinney_Lat11,Davies:2012qf}. \label{inputs}}
\end{center}
\end{table}

Among all quantities required in the UT fit, the value of $|V_{ub}|$ is the most problematic.  The determinations of $|V_{ub}|$ obtained from inclusive and exclusive semileptonic $b\to u \ell\nu \; (\ell=e,\mu)$ decays differ at the 3.3  $\sigma$ level, and may indicate the presence of underestimated uncertainties.  For the ``standard fit" presented in Sec.~\ref{sec:sf} we use a weighted average of inclusive and exclusive $|V_{ub}|$ with an error rescaled by $\sqrt{\chi^2/{\rm dof}} = 3.3$ following the PDG prescription.  In Sec.~\ref{sec:fitnovub} we also consider an alternative fit in which we omit the constraint from semileptonic $b\to u \ell\nu \; (\ell=e,\mu)$ decays.  We consider this the ``clean fit" because all remaining inputs are on excellent theoretical ground.  (Although there is also a slight disagreement between the determinations of $|V_{cb}|$ obtained from inclusive and exclusive semileptonic $b\to c \ell\nu \; (\ell=e,\mu)$, the disagreement is less than $2\sigma$ and we do not consider it a major concern.) 

\subsection{Interpretation as New Physics}
Given the presence of a tension in the CKM unitarity triangle, we can use a model-independent approach to test the compatibility of the data with various new-physics scenarios.  On general grounds, NP contributions to processes that appear at the 1-loop level in the SM ($K$, $B_d$ and $B_s$ mixing) are expected to be sizable. On the other hand, tree--level SM processes do not usually receive large corrections. An exception to this statement are contributions to $B\to \tau \nu$ ({\it e.g.} charged-Higgs exchange) and to exclusive $b\to u \ell\nu \; (\ell=e,\mu)$ decays ({\it e.g.} loop--induced effective right--handed $W-u_R-b_R$ interaction in the MSSM). New-physics contributions to $B_s$ mixing are mostly decoupled from the fit ($\Delta M_{B_s}$ does not depend on $\rho$ and $\eta$) and will not be considered here.   Given these considerations, in Sec.~\ref{sec:Sec3} we consider scenarios in which the new physics is either in kaon mixing, $B_d$-mixing, or $B\to \tau \nu$ (details of the analysis method are given in Ref.~\cite{Lunghi:2009sm}). In Sec.~\ref{sec:RH} we consider the possibility of new physics effects in exclusive $b\to u \ell\nu \; (\ell=e,\mu)$ decays.

We adopt a model-independent parametrization of new physics effects in the three observables:
\bea
|\varepsilon_K^{\rm NP}| &=& C_\varepsilon \; |\varepsilon_K^{\rm SM}| \; , \\
M_{12}^{d,{\rm NP}} &=& r_d^2 \; e^{2 i \theta_d} \; M_{12}^{d,{\rm SM}} \;, \\
{\rm BR} (B\to \tau\nu)^{\rm NP} &=&  \left(
1- \frac{\tan^2 \beta \; m_{B^+}^2}{m_{H^+}^2 (1+\epsilon_0 \tan\beta)} \right) {\rm BR} (B\to \tau\nu)^{\rm SM} \; \\
&=&r_H \;  {\rm BR} (B\to\tau\nu)^{\rm SM}  \; ,
\eea
where in the Standard Model ($C_\varepsilon, \; r_H, \; r_d )= 1$ and $\theta_d = 0$. In presence of non-vanishing contributions to $B_d$ mixing the following other observables are also affected:
\bea
S_{\psi K_S} &=& \sin 2 (\beta + \theta_d)  \; , \\
\sin (2 \alpha_{\rm eff})  &=& \sin 2 (\alpha - \theta_d)  \; , \\
X_{sd} &=& \frac{\Delta M_{B_s}}{\Delta M_{B_d}} = X_{sd}^{\rm SM} \; r_d^{-2} \; .
\eea
When considering new physics in $B_d$ mixing we allow simultaneous variations of both $\theta_d$ and $r_d$. We find that new physics in $|M_{12}^d|$ has a limited effect on the tension between the direct and indirect determinations of $\sin (2\beta)$; as a consequence, our results for $r_d$ and $\theta_d$ point to larger effects on the latter. 

Finally we interpret the constraints on the parameters $C_\varepsilon, \; r_d$, and $\theta_d$ in terms of generic new physics contributions to $\Delta S = 2$ and $\Delta B = 2$ four-fermion operators. The most general effective Hamiltonian for $B_d$--mixing can be written as~\footnote{The Hamiltonians for $B_s$-- and $K$--mixing are obtained by replacing $(d,b) \to (s,b)$ and $(d,b) \to (d,s)$, respectively.}
\bea
{\cal H}_{\rm eff} = \frac{G_F^2 m_W^2}{16 \pi^2} \left( V_{tb}^{} V_{td}^*\right)^2 \left( 
\sum_{i=1}^5 C_i  O_i  +  \sum_{i=1}^3 \tilde C_i  \tilde O_i \right) \,,
\eea
where
\begin{eqnarray}
\begin{tabular}{lcl}
$O_1 =  ( \bar d_L \gamma_\mu b_L) ( \bar d_L \gamma_\mu b_L)$
& \phantom{ciaciacia} &
$\tilde O_1  =  ( \bar d_R \gamma_\mu b_R) ( \bar d_R \gamma_\mu b_R)$
\cr
$O_2  =  ( \bar d_R  b_L) ( \bar d_R  b_L) $
& &
$\tilde O_2  =  ( \bar d_L  b_R) ( \bar d_L  b_R)$
\cr
$O_3  =  ( \bar d^{\alpha}_R  b_L^\beta ) ( \bar d^{\beta}_R  b_L^\alpha )$
& &
$\tilde O_3  =  ( \bar d^{\alpha}_L b_R^\beta ) ( \bar d^{\beta}_L  b_R^\alpha ) $
\cr
$O_4  =  ( \bar d_R  b_L) ( \bar d_L  b_R) $ 
& &
$O_5  =  ( \bar d^{\alpha}_R  b_L^\beta ) ( \bar d^{\beta}_L  b_R^\alpha )$ .
\cr
\end{tabular}
\end{eqnarray}
Within the Standard Model, only the operator $O_1$ receives a non-vanishing contribution at a high scale $\mu_H \sim m_t$. For our analysis we assume that all new physics effects can be effectively taken into account by a suitable contribution to $C_1$:
\bea
{\cal H}_{\rm eff} = \frac{G_F^2 m_W^4}{16 \pi^2} \left( V_{tb}^{} V_{td}^*\right)^2 C_1^{\rm SM} 
\left( \frac{1}{m_W^2} - \frac{e^{i\varphi}}{ \Lambda^2} \right) O_1 \; ,
\label{np}
\eea
where the minus sign has been introduced {\it a posteriori} (as we will see that the fit will point to new physics phases $\varphi \sim O(1)$). In this parametrization $\Lambda$ is the scale of some new physics model whose interactions are identical to the Standard Model with the exception of an additional arbitrary $CP$ violating phase:  
\bea
C_1 = C_1^{\rm SM} \left( 1 - e^{i\varphi} \frac{m_W^2}{\Lambda^2} \right) \;.
\label{lamphi}
\eea
%
%
\section{Unitarity Triangle Fit Results and Constraints on New Physics}
\label{sec:Sec3}
In this section we present the results obtained for the full fit and for the fits in which semileptonic decays (for the extraction of $|V_{ub}|$ and $|V_{cb}|$) are not used.  For each set of constraints we present the fitted values of the CKM parameters $\bar\rho$, $\bar\eta$ and $A$. We also show the predictions for several interesting quantities (most importantly $S_{\psi K}$ and ${\rm BR} (B\to \tau\nu)$) that we obtain after removing the corresponding direct determination from the fit.  Finally we interpret the observed discrepancies in terms of new physics in $\varepsilon_K$, $B_d$--mixing or $B\to \tau\nu$. 

\begin{figure}[t]
\begin{center}
\includegraphics[width=0.6 \linewidth]{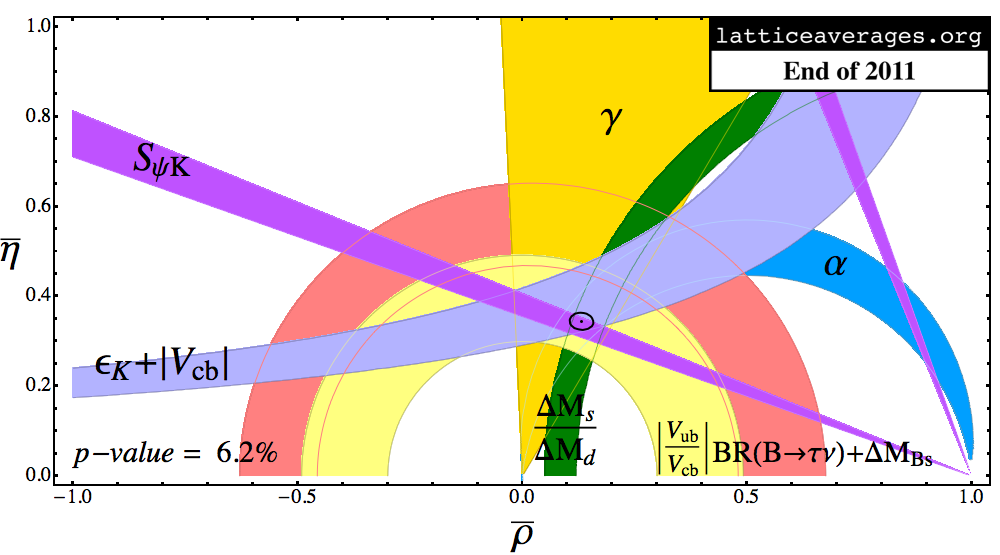}
\includegraphics[width=0.6 \linewidth]{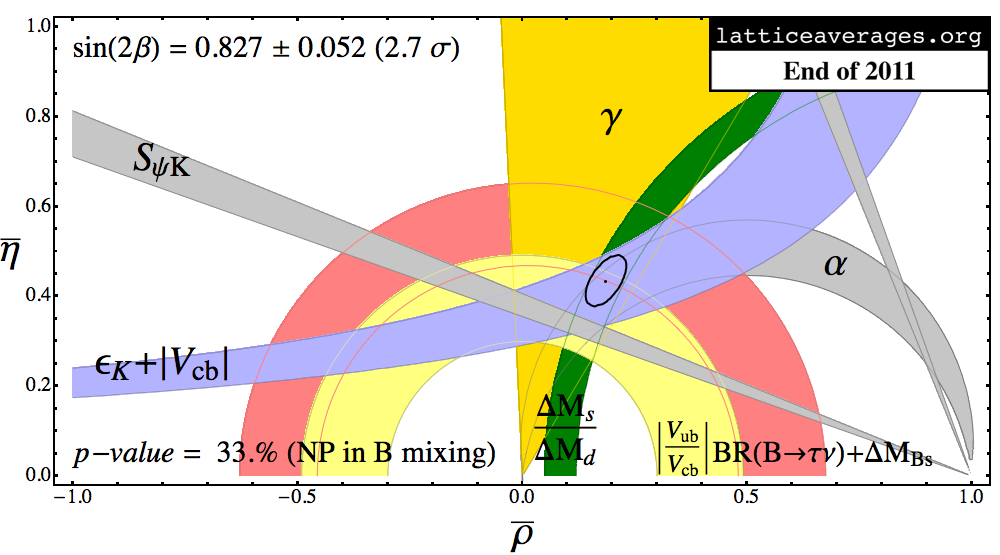}
\includegraphics[width=0.6\linewidth]{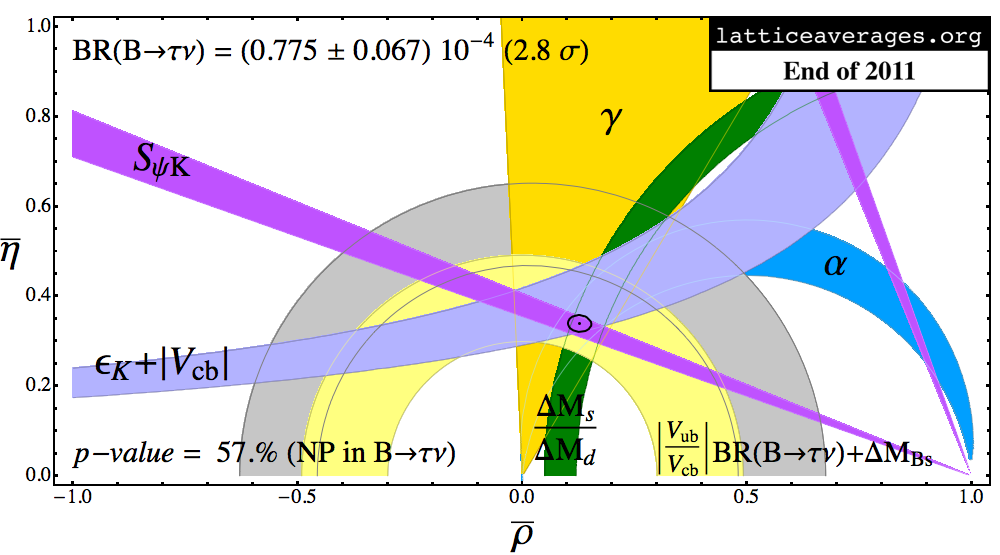}

\includegraphics[width=0.35 \linewidth]{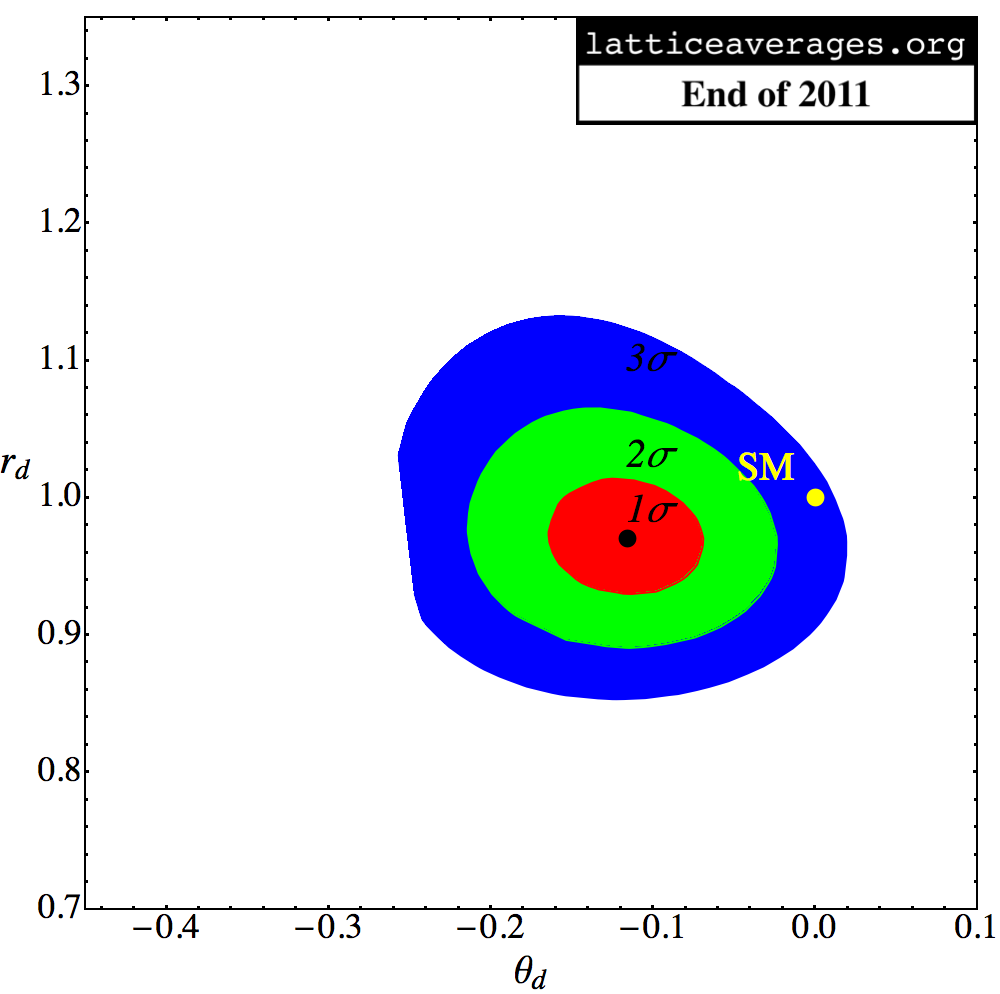}
\includegraphics[width=0.35 \linewidth]{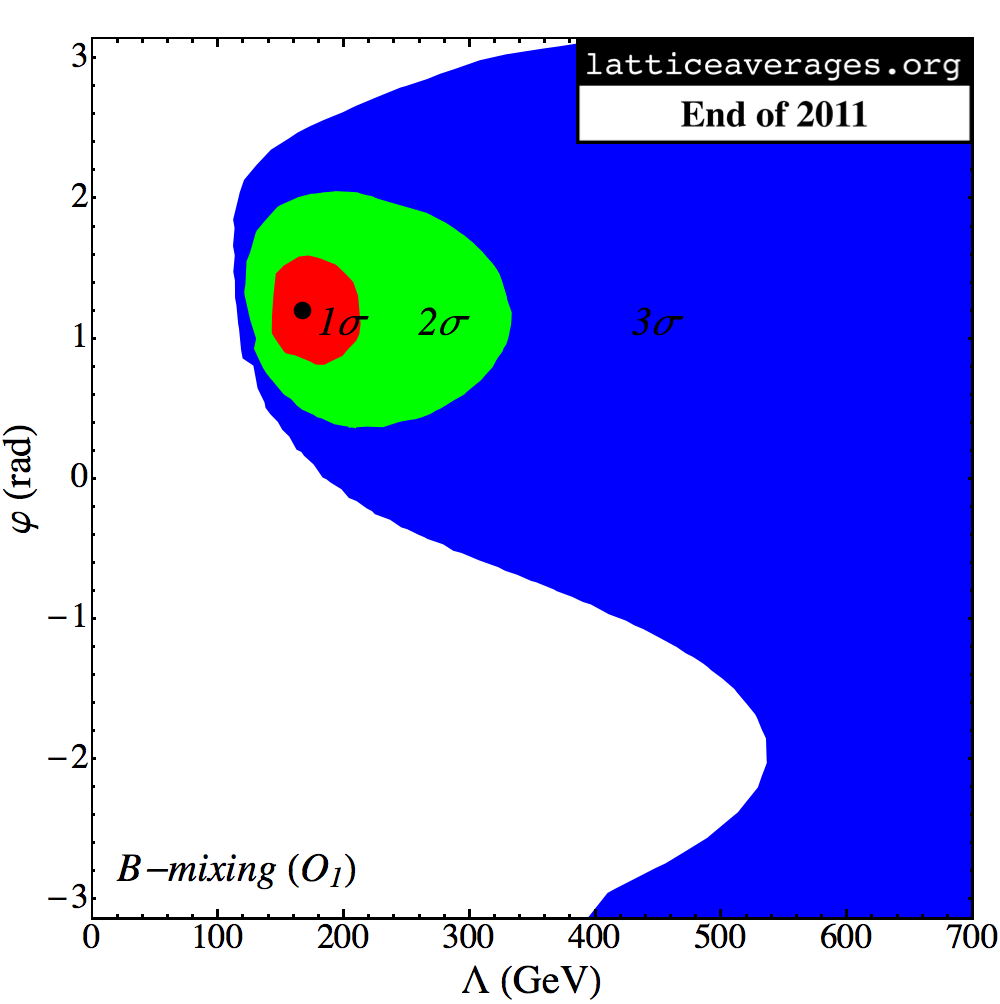}
\caption{Unitarity triangle fit with all constraints included. Quantities that are not used to generate the black contour are grayed out. \label{fig:utfit-tot}}
\end{center}
\end{figure}

\begin{figure}[t]
\begin{center}
\includegraphics[width= 0.6 \linewidth]{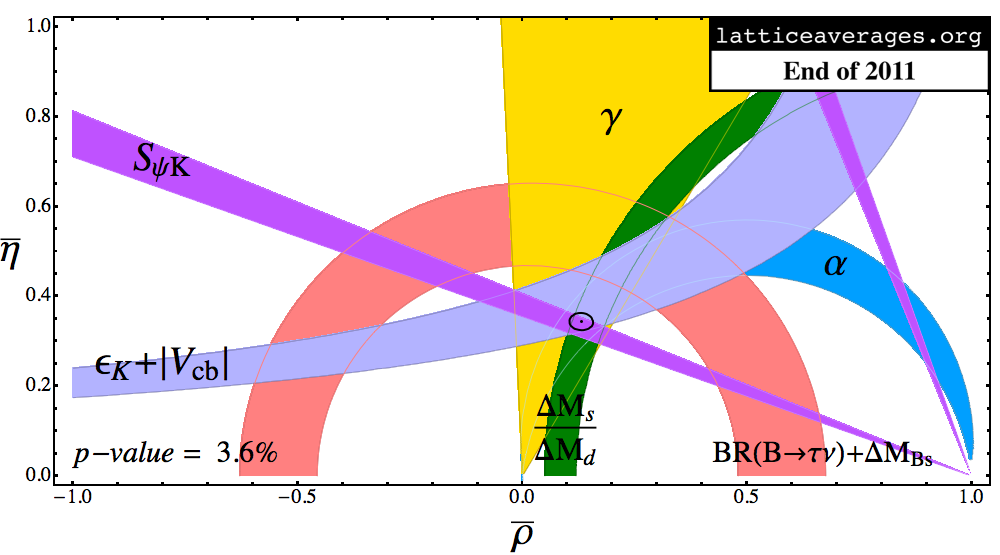}
\includegraphics[width= 0.6 \linewidth]{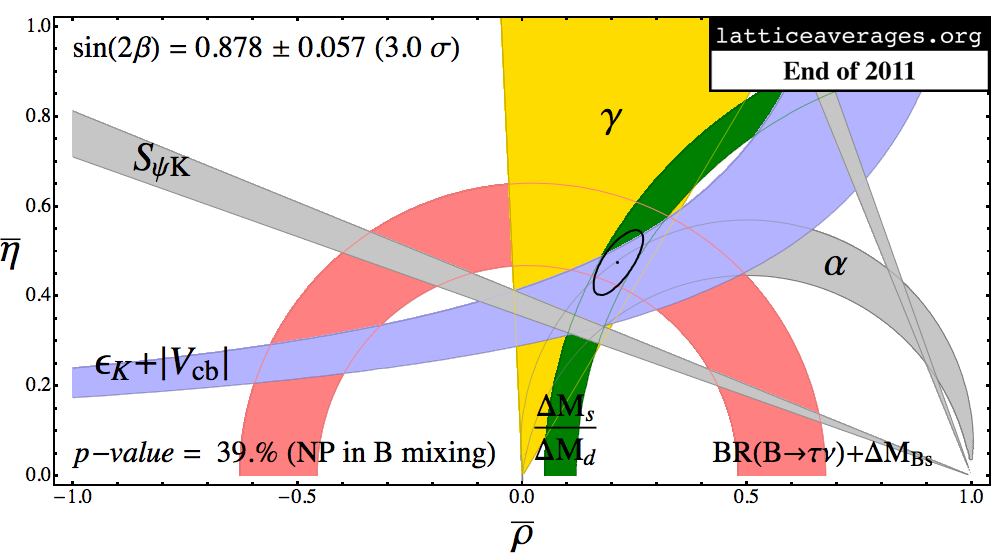}
\includegraphics[width= 0.6 \linewidth]{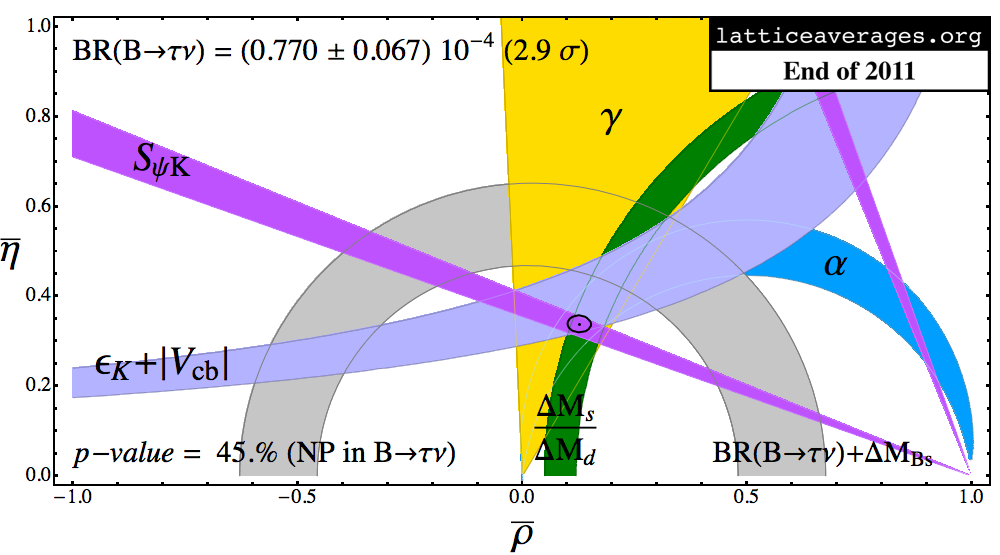}

\includegraphics[width=0.35 \linewidth]{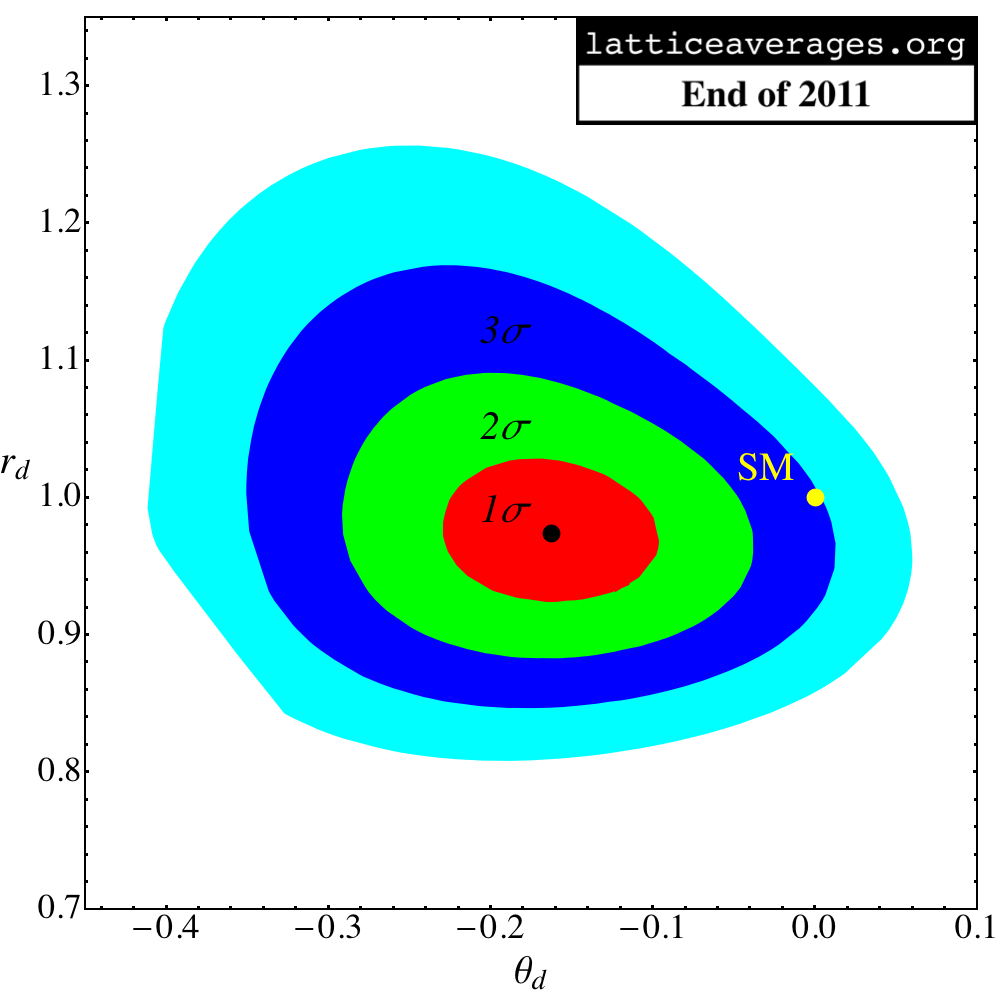}
\includegraphics[width=0.35 \linewidth]{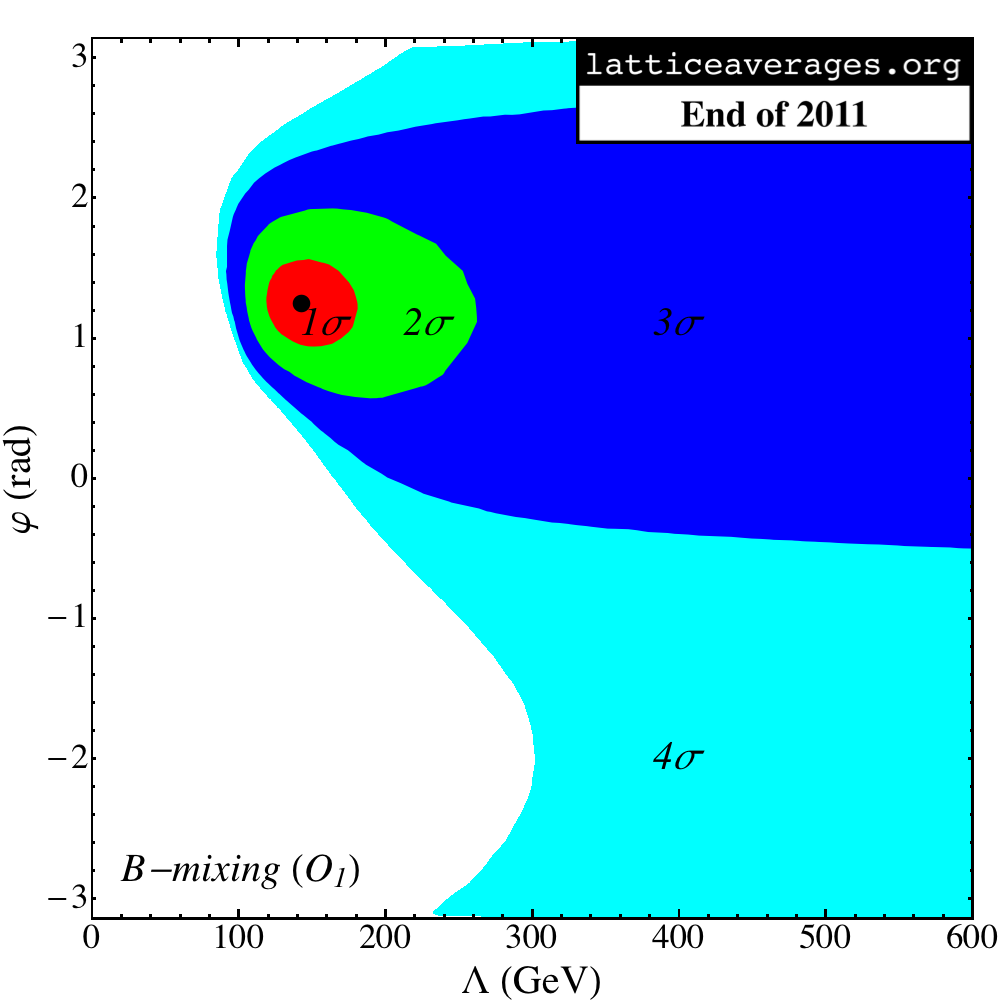}
\caption{Unitarity triangle fit without $|V_{ub}|$. Quantities that are not used to generate the black contour are grayed out.\label{fig:utfit-novub}}
\end{center}
\end{figure}

\begin{figure}[t]
\begin{center}
\includegraphics[width=0.6 \linewidth]{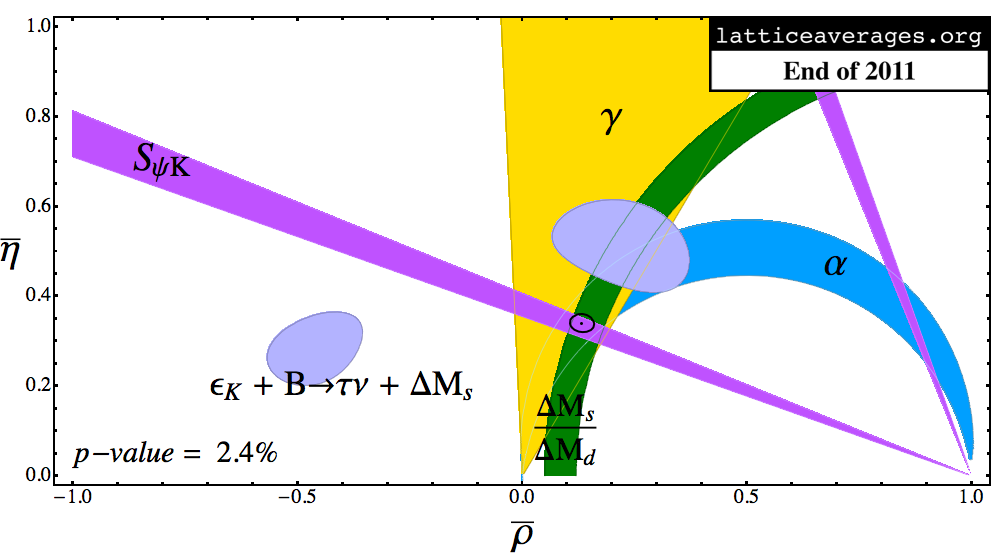} 
\includegraphics[width=0.6 \linewidth]{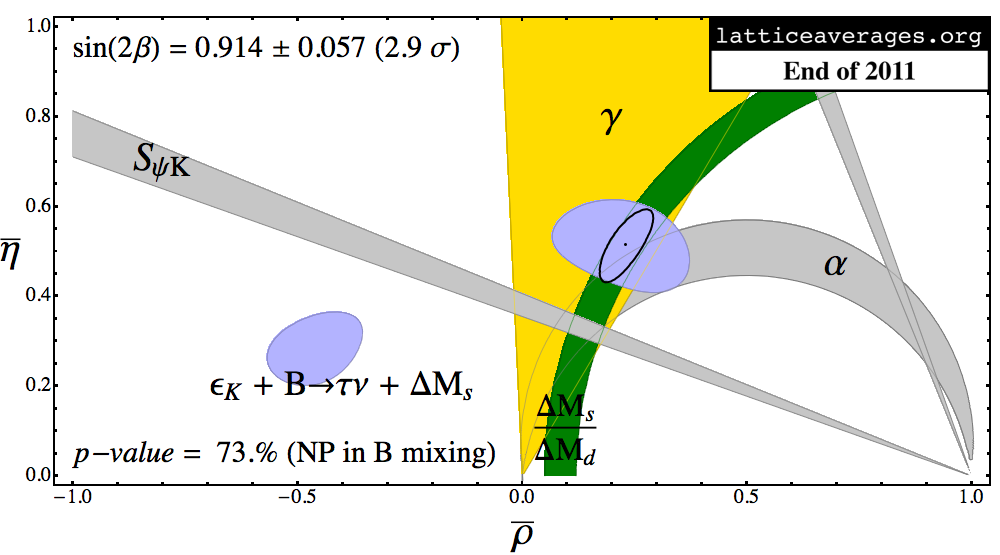} 
\includegraphics[width=0.6 \linewidth]{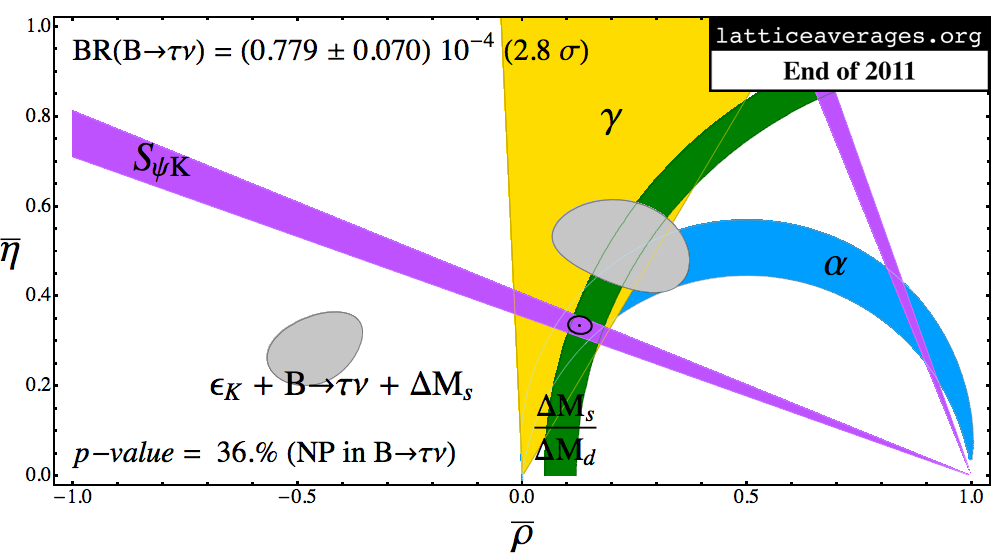} 

\includegraphics[width=0.35 \linewidth]{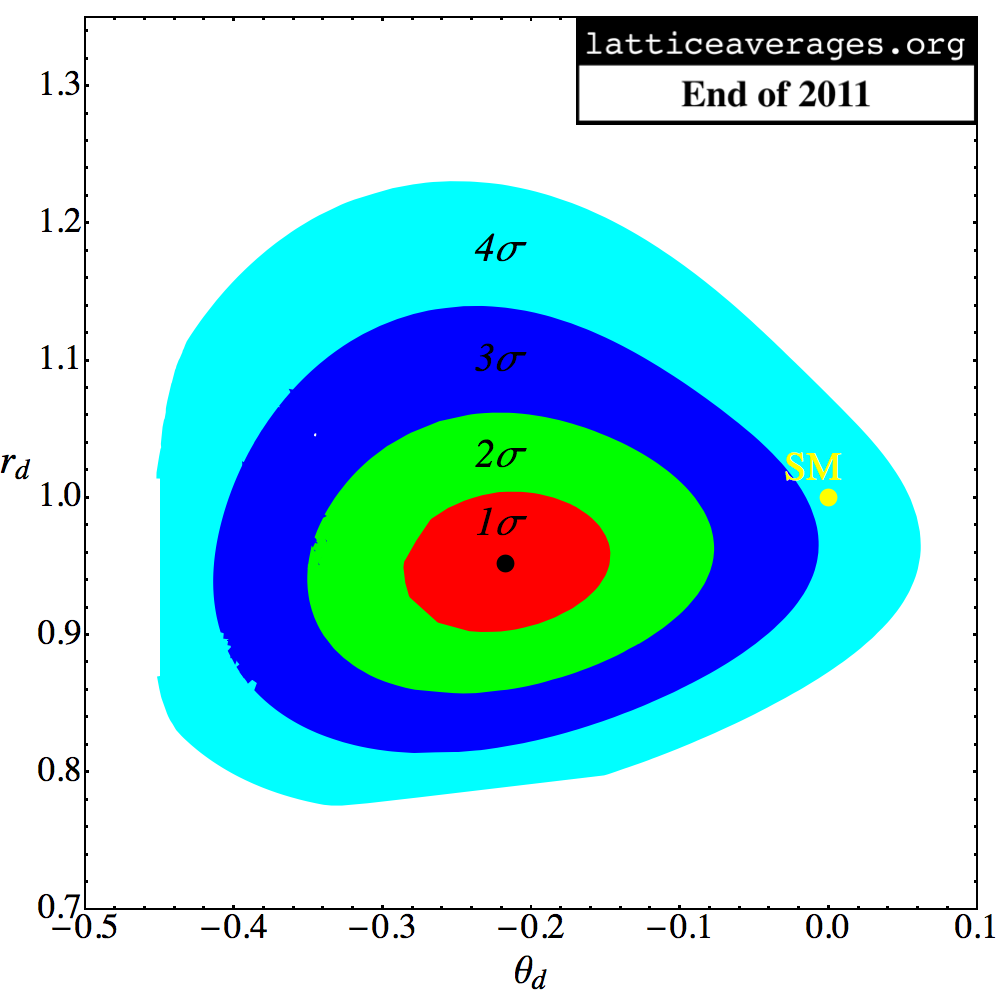}
\includegraphics[width=0.35 \linewidth]{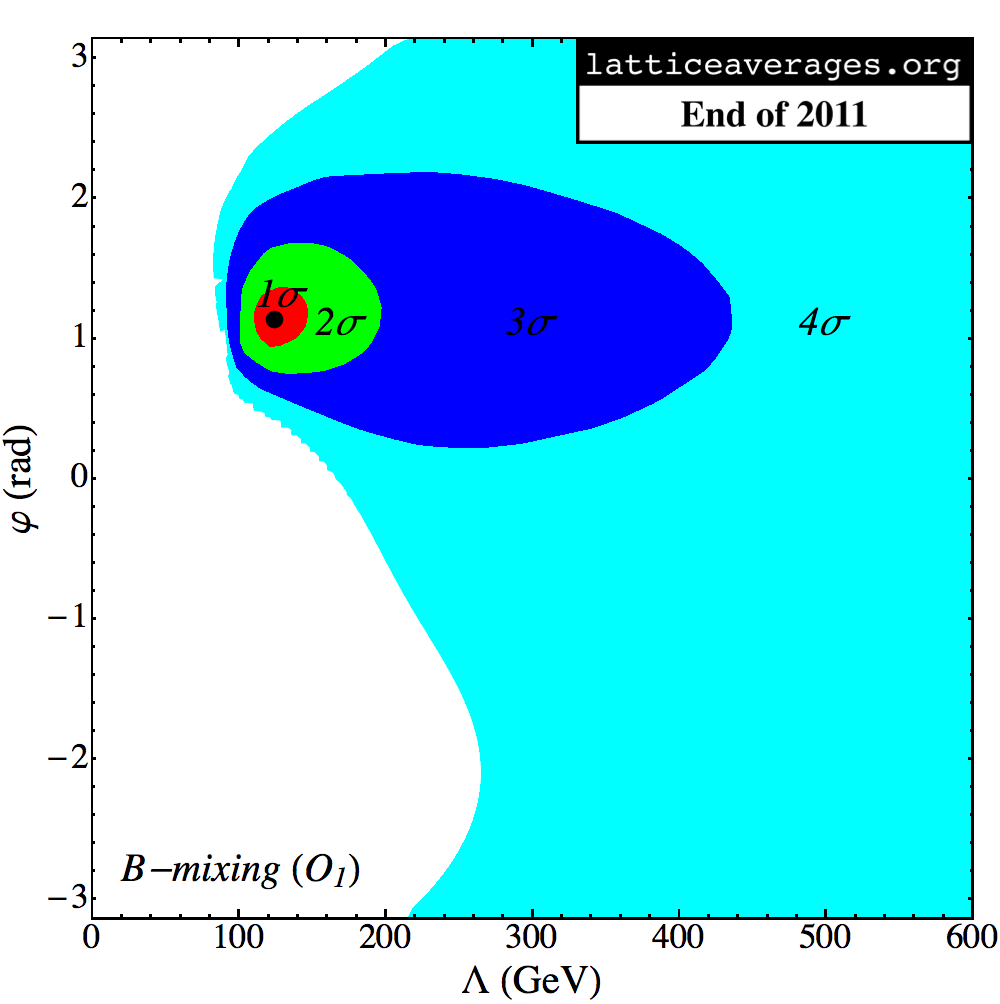}
\caption{Unitarity triangle fit without $|V_{ub}|$ and $|V_{cb}|$. Quantities that are not used to generate the black contour are grayed out.\label{fig:utfit-novqb}}
\end{center}
\end{figure}

In the upper panels of Figs.~\ref{fig:utfit-tot}--\ref{fig:utfit-novqb}, we show the global CKM unitarity triangle fit for the three sets of inputs that we consider (complete fit, no $V_{ub}$ fit, no $V_{qb}$ fit). In each figure, the black contours and $p$--values in the top, middle and bottom panels correspond to the complete fit, the fit with a new phase in $B$ mixing ({\it i.e.} without using $S_{\psi K}$ and $\alpha$) and the fit with new physics in $B\to \tau \nu$ ({\it i.e.} without using ${\rm BR} (B\to\tau\nu)$), respectively. For fits with a new phase in $B$ mixing we show the fit predictions for $\sin(2\beta)$, while for the fits with new physics in $B\to \tau \nu$, we show the fit predictions for ${\rm BR} (B\to\tau\nu)$. (Note that the individual contours in these figures never use the same constraint twice: in particular, the $B\to\tau\nu$ allowed area is obtained by using $\Delta M_{B_s}$ instead of the direct determination of $|V_{cb}|$.)

The lower panels of Figs.~\ref{fig:utfit-tot}--\ref{fig:utfit-novqb} show the interpretation of the tensions highlighted in the various fits in terms of possible new-physics scenarios. In the first panel of each figure we present the result of the two-dimensional fit in the $(\theta_d,r_d)$ plane and in the second panel we map this allowed region onto the $(\Lambda,\varphi)$ plane (see Eq.~(\ref{lamphi})) under the assumption of new physics in $O_1$ only. We do not show the corresponding plots for new physics in $K$ mixing because these contributions cannot relieve the tension in the fit (as can be seen by the poor $p$-values in Eqs.~(\ref{cepsilonTOT}), (\ref{cepsilon}) and (\ref{cepsilonNOVQB})).
\subsection{Standard Fit}
\label{sec:sf}
\noindent We include constraints from $\varepsilon_K$, $\Delta M_{B_d}$, $\Delta M_{B_s}$, $\alpha$, $S_{\psi K}$, $\gamma$, ${\rm BR} (B\to\tau\nu)$, $|V_{cb}|$ and $|V_{ub}|$. The overall $p$-value of the Standard-Model fit is $ p = 6.2\% $ and the results of the fit are
\beq
\bar \rho  =  0.131 \pm 0.018
 \quad\quad 
\bar \eta  =  0.344 \pm 0.013
 \quad\quad 
A  =  0.818 \pm 0.014
  
\; .
\eeq
The predictions from all other information when the direct determination of the quantity is removed from the fit are
\begin{align}
& |V_{ub}|  =  (3.53 \pm 0.13 ) \; \times 10^{-3} \quad (0.4\; \sigma)
 \\
& S_{\psi K} = 0.827 \pm 0.052 \quad (2.7\; \sigma)
 \\
& |V_{cb}| = (42.1 \pm 1.0 ) \; \times 10^{-3} \quad (0.8\; \sigma)
 \\
& \hat B_K = 0.880 \pm 0.11 \quad (1.0\; \sigma)
 \\
& f_{B_d} \sqrt{\hat B_d}  =  (209.3 \pm 4.8 ) \; {\rm MeV}   \quad (1.0\; \sigma)
 \\
& {\rm BR} (B\to\tau\nu)  =  (0.775 \pm 0.067 ) \; \times 10^{-4} \quad (2.8\; \sigma)
 \\
& f_{B_d} = (280. \pm 28. ) \; {\rm MeV} \quad (3.2\; \sigma)
  
\end{align}
where we indicate the deviation from the corresponding direct determination in parentheses.
The interpretation of the above discrepancies in terms of new physics in $K$--mixing, $B_d$--mixing and $B\to \tau \nu$ yields
\begin{align}
& \hphantom{5\;\;} C_\varepsilon  =  1.15 \pm 0.14 \quad\quad ( 1.0\; \sigma ,\;  p = 0.053)

  \label{cepsilonTOT} \\ 
&\begin{cases}
\theta_d  =  - (6.7 \pm 2.8)^{\rm o}
   \\
r_d  =  0.97 \pm 0.042
   \\ 
\end{cases}  ( 2.5\; \sigma ,\;  p = 0.33
 )\\
&\hphantom{5\;\;}  r_H  =  2.18 \pm 0.44 \quad\quad ( 2.9\; \sigma ,\;  p = 0.57)
 
\; .
\end{align}
In Figure~\ref{fig:utfit-tot} we summarize these results and present the two dimensional $(\theta_d,r_d)$ allowed regions and the interpretation of these results in terms of a minimal flavor violating new-physics scale.
\subsection{Fit Without $|V_{ub}|$}
\label{sec:fitnovub}
\noindent We include constraints from $\varepsilon_K$, $\Delta M_{B_d}$, $\Delta M_{B_s}$, $\alpha$, $S_{\psi K}$, $\gamma$, ${\rm BR} (B\to\tau\nu)$ and $|V_{cb}|$. The overall $p$-value of the Standard-Model fit is $ p = 3.6\% $ and the results of the fit are
\beq
\bar \rho  =  0.131 \pm 0.018
 \quad\quad 
\bar \eta  =  0.344 \pm 0.013
 \quad\quad 
A  =  0.818 \pm 0.014
  
\; .
\eeq
The predictions from all other information when the direct determination of the quantity is removed from the fit are
\begin{align}
& |V_{ub}|  =  (3.53 \pm 0.13 ) \; \times 10^{-3} \quad (0.4\; \sigma)
 \\
& S_{\psi K} = 0.878 \pm 0.057 \quad (3.0\; \sigma)
 \\
& |V_{cb}| = (42.1 \pm 1.0 ) \; \times 10^{-3} \quad (0.8\; \sigma)
 \\
& \hat B_K = 0.888 \pm 0.11 \quad (1.1\; \sigma)
 \\
& f_{B_d} \sqrt{\hat B_d}  =  (209.3 \pm 4.8 ) \; {\rm MeV}  \quad (1.0\; \sigma)
 \\
& {\rm BR} (B\to\tau\nu)  =  (0.770 \pm 0.067 ) \; \times 10^{-4} \quad (2.9\; \sigma)
 \\
& f_{B_d} = (281. \pm 28. ) \; {\rm MeV} \quad (3.2\; \sigma)
  
\end{align}
where we indicate the deviation from the corresponding direct determination in parentheses.
The interpretation of the above discrepancies in terms of new physics in $K$--mixing, $B_d$--mixing and $B\to \tau \nu$ yields
\begin{align}
& \hphantom{5\;\;} C_\varepsilon  =  1.16 \pm 0.14 \quad\quad ( 1.1\; \sigma ,\;  p = 0.031)
     \label{cepsilon}  \\ 
&\begin{cases}
\theta_d  =  - (9.3 \pm 3.5)^{\rm o}
   \\
r_d  =  0.98 \pm 0.046
   \\ 
\end{cases}  ( 2.8\; \sigma ,\;  p = 0.39
 )\\
&\hphantom{5\;\;}  r_H  =  2.20 \pm 0.45 \quad\quad ( 2.9\; \sigma ,\;  p = 0.45)
    \label{rh} 
\; .
\end{align}
In Figure~\ref{fig:utfit-novub} we summarize these results and present the two dimensional $(\theta_d,r_d)$ allowed regions and the interpretation of these results in terms of a minimal flavor violating new-physics scale.
\subsection{Fit Without $|V_{ub}|$ and $|V_{cb}|$}
\label{sec:fitnovqb}
\noindent   The strategy for removing $|V_{cb}|$ by combining the constraints from $\varepsilon_K$, $\Delta M_{B_s}$, and ${\rm BR} (B\to \tau \nu)$ was proposed and is described in detail in Ref.~\cite{Lunghi:2009ke}. We include constraints from $\varepsilon_K$, $\Delta M_{B_d}$, $\Delta M_{B_s}$, $\alpha$, $S_{\psi K}$, $\gamma$ and ${\rm BR} (B\to\tau\nu)$. The overall $p$-value of the Standard-Model fit is $  p = 2.4\%  $ and the results of the fit are
\beq
\bar \rho  =  0.132 \pm 0.018
 \quad\quad 
\bar \eta  =  0.341 \pm 0.013
 \quad\quad 
A  =  0.829 \pm 0.020
  
\; .
\eeq
The predictions from all other information when the direct determination of the quantity is removed from the fit are
\begin{align}
& |V_{ub}|  =  (3.55 \pm 0.13 ) \; \times 10^{-3} \quad (0.3\; \sigma)
 \\
& S_{\psi K} = 0.914 \pm 0.057 \quad (2.9\; \sigma)
 \\
& |V_{cb}| = (42.1 \pm 1.0 ) \; \times 10^{-3} \quad (0.8\; \sigma)
 \\
& \hat B_K = 0.918 \pm 0.18 \quad (0.8\; \sigma)
 \\
& f_{B_d} \sqrt{\hat B_d}  =  (207.2 \pm 5.4 ) \; {\rm MeV}  \quad (1.1\; \sigma)
 \\
& {\rm BR} (B\to\tau\nu)  =  (0.779 \pm 0.070 ) \; \times 10^{-4} \quad (2.8\; \sigma)
 \\
& f_{B_d} = (280. \pm 28. ) \; {\rm MeV} \quad (3.1\; \sigma)
  
\end{align}
where we indicate the deviation from the corresponding direct determination in parentheses.
The interpretation of the above discrepancies in terms of new physics in $K$--mixing, $B_d$--mixing and $B\to \tau \nu$ yields
\begin{align}
& \hphantom{5\;\;} C_\varepsilon  =  1.20 \pm 0.24 \quad\quad ( 0.78\; \sigma ,\;  p = 0.014)
  
   \label{cepsilonNOVQB}\\ 
&\begin{cases}
\theta_d  =  - (12. \pm 3.8)^{\rm o}
   \\
r_d  =  0.95 \pm 0.048
   \\ 
\end{cases}  ( 3.1\; \sigma ,\;  p = 0.73
 )\\
&\hphantom{5\;\;}  r_H  =  2.18 \pm 0.44 \quad\quad ( 2.8\; \sigma ,\;  p = 0.36)
 
\; .
\end{align}
In Figure~\ref{fig:utfit-novqb} we summarize these results and present the two dimensional $(\theta_d,r_d)$ allowed regions and the interpretation of these results in terms of a minimal flavor violating new-physics scale.
\section{Separate Treatment of Inclusive and Exclusive $|V_{ub}|$}
\label{sec:RH}

\begin{figure}[H]
\begin{center}
\includegraphics[width=0.6 \linewidth]{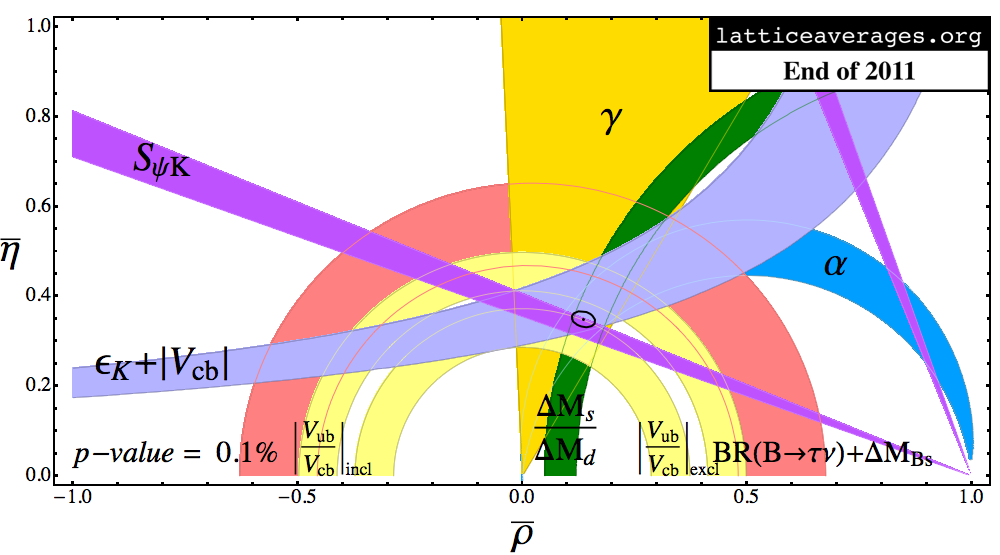}

\includegraphics[width=0.4 \linewidth]{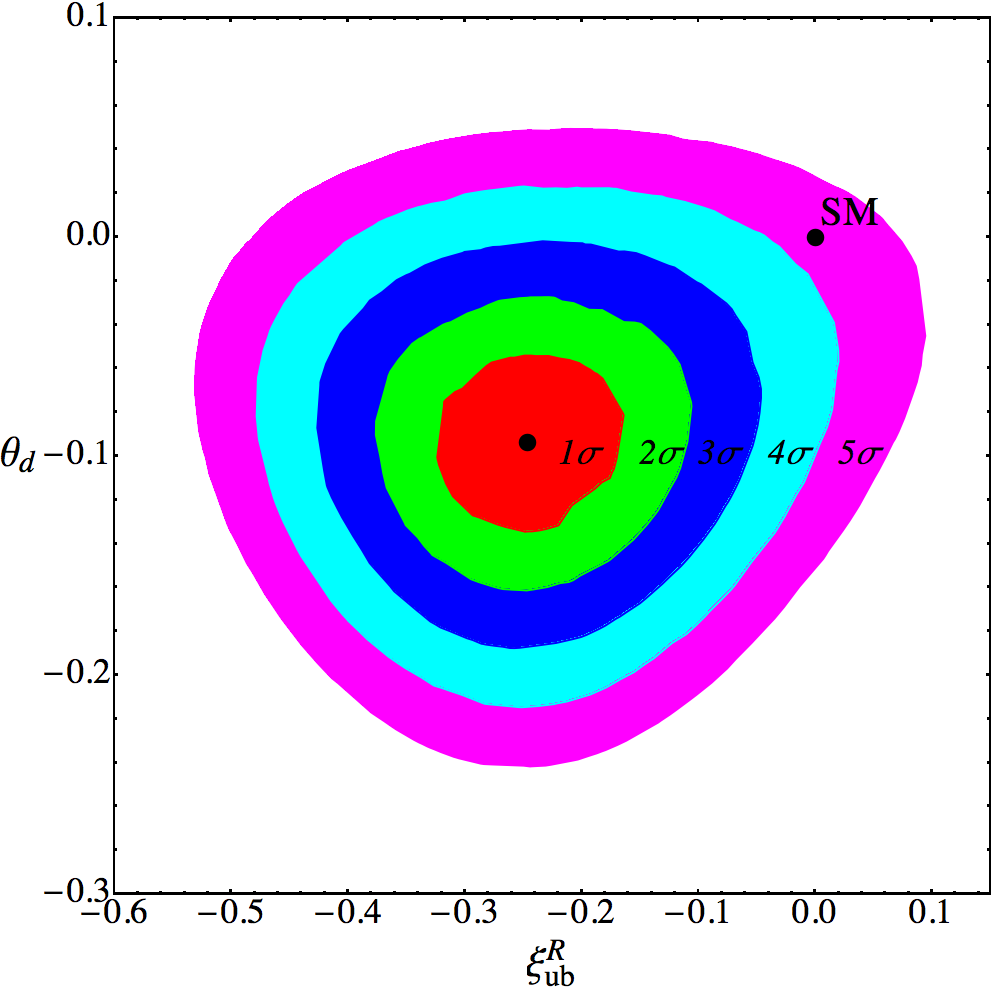}
\caption{Unitarity triangle fit with all constraints included (inclusive and exclusive $V_{ub}$ are included separately in the fit.). \label{fig:utfit-tot2}}
\end{center}
\end{figure}

In this section we approach the full fit to the unitarity triangle from a different perspective: instead of averaging the extractions of $|V_{ub}|$ from inclusive and exclusive semileptonic $b\to u\ell\nu \; (\ell = e,\mu)$ decays (and inflating the resulting error), we take the $3.3\sigma$ disagreement between the two determinations at face value. The fit therefore includes constraints from $\varepsilon_K$, $\Delta M_{B_d}$, $\Delta M_{B_s}$, $\alpha$, $S_{\psi K}$, $\gamma$, ${\rm BR} (B\to\tau\nu)$, $|V_{cb}|$ and $|V_{ub}|^{\rm excl}$ and $|V_{ub}|^{\rm incl}$. The overall $p$-value of the fit is now very small ($p = 0.1\% $) and the results we obtain are
\beq
\bar \rho  =  0.136 \pm 0.017
 \quad\quad 
\bar \eta  =  0.349 \pm 0.012
 \quad\quad 
A  =  0.820 \pm 0.014
  
\; .
\eeq
The predictions from all other information when the direct determination of the quantity is removed from the fit are
\begin{align}
& |V_{ub}|  =  (3.53 \pm 0.13 ) \; \times 10^{-3} \quad (0.4\; \sigma)
 \\
& S_{\psi K} = 0.750 \pm 0.029 \quad (2.3\; \sigma)
 \\
& |V_{cb}| = (42.25 \pm 0.98 ) \; \times 10^{-3} \quad (0.9\; \sigma)
 \\
& \hat B_K = 0.838 \pm 0.097 \quad (0.7\; \sigma)
 \\
& f_{B_d} \sqrt{\hat B_d}  =  (209.3 \pm 4.9 ) \; {\rm MeV}   \quad (1.0\; \sigma)
 \\
& {\rm BR} (B\to\tau\nu)  =  (0.813 \pm 0.063 ) \; \times 10^{-4} \quad (2.7\; \sigma)
 \\
& f_{B_d} = (273. \pm 27. ) \; {\rm MeV} \quad (3.1\; \sigma)
  
\end{align}
where we indicate the deviation from the corresponding direct determination in parentheses.
The interpretation of the above discrepancies in terms of new physics in $K$--mixing, $B_d$--mixing and $B\to \tau \nu$ yields
\begin{align}
& \hphantom{5\;\;} C_\varepsilon  =  1.09 \pm 0.13 \quad\quad ( 0.68\; \sigma ,\;  p = 0.00063)

  \label{cepsilonTOT-complete2} \\ 
&\begin{cases}
\theta_d  =  - (3.4 \pm 1.5)^{\rm o}
   \\
r_d  =  0.96 \pm 0.038
   \\ 
\end{cases}  ( 2.3\; \sigma ,\;  p = 0.0032
 )\\
&\hphantom{5\;\;}  r_H  =  2.08 \pm 0.41 \quad\quad ( 2.7\; \sigma ,\;  p = 0.011)
 
\; .
\end{align}
Figure~\ref{fig:utfit-tot2} summarizes these results. Note that none of the new physics scenarios we consider is able to lift the overall $p$-value of the fit: the tension between the inclusive and exclusive determination of $|V_{ub}|$, if taken at face value, can only be addressed by dedicated new-physics contributions. 

Alleviating the discrepancy between inclusive and exclusive $|V_{ub}|$ requires the introduction of interactions whose impact on exclusive $B\to X_u \ell \nu$ decays is much larger than in inclusive ones.  The introduction of a right--handed effective $\bar u_R \Wslash b_R$ coupling offers the most elegant solution of the ``$V_{ub}$ puzzle'' (see for instance Refs.~\cite{Chen:2008se,Crivellin:2009sd,Buras:2010pz,Blanke:2011ry}). In this scenario we have:
\begin{eqnarray}
V_{ub} \; \bar u_L \Wslash b_L \Longrightarrow V_{ub} \; \left( \bar u_L \Wslash b_L + \xi_{ub}^R \;  \bar u_R \Wslash b_R \right) \; .
\end{eqnarray}
The effective parameter $\xi_{ub}^R$ affects all $b\to u \ell\nu \; (\ell = e,\mu,\tau)$ transitions:
\begin{align}
\left| V_{ub} \right|_{\rm incl} &\Longrightarrow \sqrt{1+ \left| \xi_{ub}^R\right|^2} \; \left| V_{ub} \right| \; , \label{eq1} \\
\left| V_{ub} \right|_{\rm excl} &\Longrightarrow \left| 1+  \xi_{ub}^R \right| \; \left| V_{ub} \right| \; ,\\
{\rm BR} (B\to \tau\nu) &\Longrightarrow  \left| 1-  \xi_{ub}^R \right|^2 \; {\rm BR} (B\to \tau\nu) \; . \label{eq3}
\end{align}
The result of the fit to the unitarity triangle in which we allow $\xi_{ub}^R$, $\theta_d$ and $r_d$ to vary simultaneously\footnote{Although the introduction of right-handed $b\to u \ell\nu$ currents can resolve the more than 3$\sigma$ tension between $|V_{ub}|_{\rm incl}$, $|V_{ub}|_{\rm excl}$, and $B\to\tau\nu$, additional new physics in $B_d$-mixing is needed to bring the global CKM unitarity-triangle fit into complete agreement.  This is because, under the assumptions in Eqs.~(\ref{eq1})--(\ref{eq3}), inclusive $|V_{ub}|$ is affected minimally by new physics, so the tension between inclusive $|V_{ub}|$ and $\sin(2\beta)$ remains.  The tenth entry in the upper panel of Fig.~\ref{fig:tabsin2beta} shows that even if exclusive $|V_{ub}|$ and $B\to \tau\nu$ are removed from the fit, there is still a 3.4$\sigma$ discrepancy.} yields
\begin{align}
\xi_{ub}^R &= -0.245 \pm 0.055 \quad\quad\quad (4.0 \sigma) \; ,\\
\theta_d &= - (4.8 \pm 1.5)^{\rm o} \quad\quad\quad\quad (3.2 \sigma) \; ,\\
r_d &= 0.978 \pm 0.041 \quad\quad\quad\quad (0.5 \sigma) \; .
\end{align}
In the lower panel of Fig.~\ref{fig:utfit-tot2} we show the two--dimensional allowed regions in the $[\xi_{ub}^R,\theta_d]$ plane.  From the direct inspection of Fig.~\ref{fig:utfit-tot2}, we see a clear correlation between $\xi_{ub}^R$ and $\theta_d$ (the allowed region is not a simple ellipsis with axes parallel to the coordinate axes). The origin of this correlation is two-fold. The inclusion of the rest of the fit allows for a more precise determination of $|V_{ub}|$ and, through Eqs.~(\ref{eq1})--(\ref{eq3}), of $\xi_{ub}^R$. The value of $\xi_{ub}^R$ that comes from Eqs.~(\ref{eq1})--(\ref{eq3}), however, implies a value of $|V_{ub}|$ that is not optimal to lift the residual tension in $\sin(2\beta)$.  Finally we note that the impact of a $3\%$ determination of ${\rm BR} (B\to\tau\nu)$ would be to reduce by a factor of two the longer axis in Fig.~\ref{fig:utfit-tot2}, while leaving the other axis unchanged.

\section{Future Expectations}

In the near future, we expect improved lattice-QCD calculations to impact the CKM unitarity-triangle fit in several ways.  Now that the uncertainty in the neutral kaon mixing matrix element $B_K$ is below 2\%, the constraint on the UT from $\epsilon_K$ is limited by the uncertainty in the Wolfenstein parameter $A$ (recall that $\varepsilon_K \propto A^4$).  Currently $b\to c\ell\nu$ decays offer the best determination of $A = |V_{cb}|/\lambda^2$ with an uncertainty of about $2\%$.    Residual theoretical uncertainties on the extraction of $|V_{cb}|$ from inclusive $B \to X_c \ell\nu$ decays will make it very hard to further reduce the error in $|V_{cb}|$ obtained from this approach.  In the next few years, however, lattice-QCD calculations of the $B\to D \ell\nu$ and $B \to D^* \ell\nu$ form factors at nonzero recoil should allow the uncertainties in direct determinations of $|V_{cb}|$ to approach $\gtapprox 1\%$.  

An alternative extraction of $A$ is also possible via $\Delta M_{B_s}$ ( $A \propto \Delta M_{B_s} / (f_{B_s} \hat B_s^{1/2})$).  (In fact, for the first time there are now at least two $N_f = 2+1$ determinations of the quantities $\xi$, $f_{B_s} \sqrt{\hat B_s}$ and $f_B$.)  Although this approach is presently limited by a lattice-QCD uncertainty of about $6\%$, future improvements in lattice-QCD calculations of  will push the latter uncertainty to the $1\%$ level, making this indirect determination of $A$ competitive with the direct extraction from semileptonic decays.  In order to study the potential of this indirect method we consider the impact of a reduced uncertainty on $f_{B_s} \hat B_s^{1/2}$, in conjunction with the expected Belle II/SuperB determinations of ${\rm BR}(B\to \tau\nu)$ with 50 ${\rm ab}^{-1}$, in our ``no $V_{qb}$'' scenario. The results are summarized in Table~\ref{tab:superb}.
\begin{table}
\begin{eqnarray}
 \begin{array}{|cc|c|ccc|} \hline
  \delta _{\tau } & \delta _s & p_{\text{SM}} & \theta _d\pm \delta \theta _d & p_{\text{$\theta$d}} & \delta \theta _d/\theta _d \\\hline
 \text{18$\%$} & \text{4.8$\%$}  & \text{2.4$\%$} & -12.4\pm 3.8 & \text{73.$\%$} & \text{3.1$\sigma $} \\
 \text{18$\%$} & \text{2.5$\%$} & \text{0.5$\%$} & -11.7\pm3.5 & \text{67.$\%$} & \text{3.5$\sigma $} \\
 \text{18$\%$} & \text{1$\%$} &  \text{0.09$\%$} & -11.4\pm 3.3& \text{64.$\%$} & \text{3.9$\sigma $} \\\hline
 \text{10$\%$} & \text{4.8$\%$}  & \text{2e-3$\%$} & -13.1\pm2.5 & \text{72.$\%$} & \text{5.1$\sigma $} \\
 \text{3$\%$} & \text{4.8$\%$}  & \text{5e-12$\%$} & -13.4\pm1.7 & \text{71.$\%$} & \text{7.2$\sigma $} \\\hline
 \text{10$\%$} & \text{2.5$\%$} & \text{2e-4$\%$} & -12.6\pm2.3 & \text{64.$\%$} & \text{5.5$\sigma $} \\
 \text{10$\%$} & \text{1$\%$}  & \text{2e-5$\%$} & -12.3\pm 2.2 & \text{59.$\%$} & \text{5.8$\sigma $} \\
 \text{3$\%$} & \text{2.5$\%$} & \text{1e-14$\%$} & -13.0\pm 1.5 & \text{62.$\%$} & \text{8.9$\sigma $} \\
 \text{3$\%$} & \text{1$\%$}  & \text{2e-17$\%$} & -12.7\pm 1.4 & \text{57.$\%$} & \text{9.5$\sigma $} \\\hline
\end{array}
\nonumber
\end{eqnarray}
\caption{Impact of improved determinations of $f_{B_s} \hat B_s^{1/2}$ and ${\rm BR} (B\to\tau\nu)$ on the no-$V_{qb}$ fit. We define $\delta_s = \delta\left[ f_{B_s} \hat B_s^{1/2}\right]$ and $\delta_\tau = \delta \left[{\rm BR} (B\to \tau\nu)\right]$.  \label{tab:superb}}
\end{table}

\begin{figure}[H]
\begin{center}
\includegraphics[width=0.6 \linewidth]{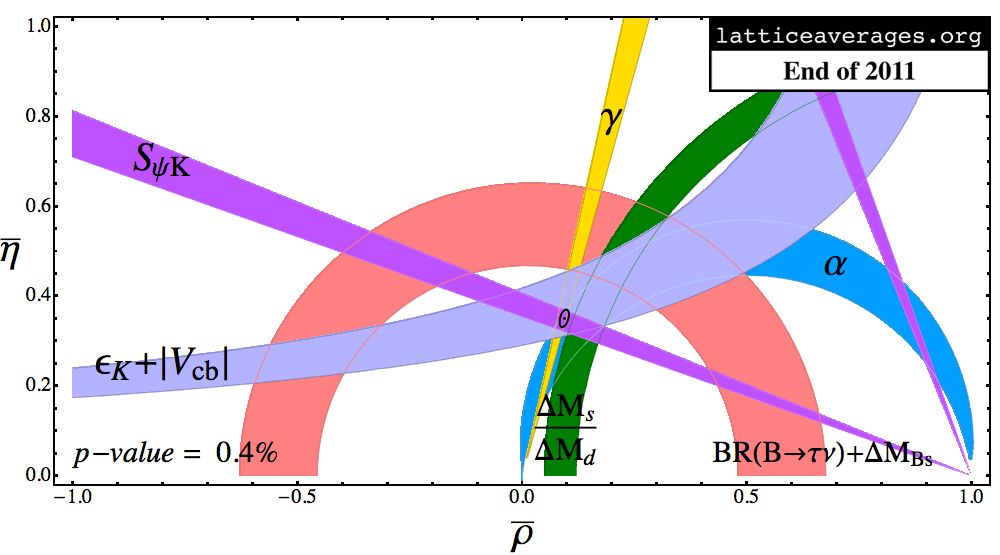}
\caption{Unitarity triangle fit without $|V_{ub}|$. We adopt a $1\%$ uncertainty on the determination of $\gamma$. \label{fig:utfitsmallgamma}}
\end{center}
\end{figure}

Finally, in Fig.~\ref{fig:utfitsmallgamma} we show the impact of a much improved determination of the angle $\gamma$. The super flavor factories (Belle II and SuperB) and BES III should be able to push the uncertainty on $\gamma$ below $1^{\rm o}$.
\section{Discussion}
\begin{figure}[H]
\begin{center}
\includegraphics[width=0.65 \linewidth]{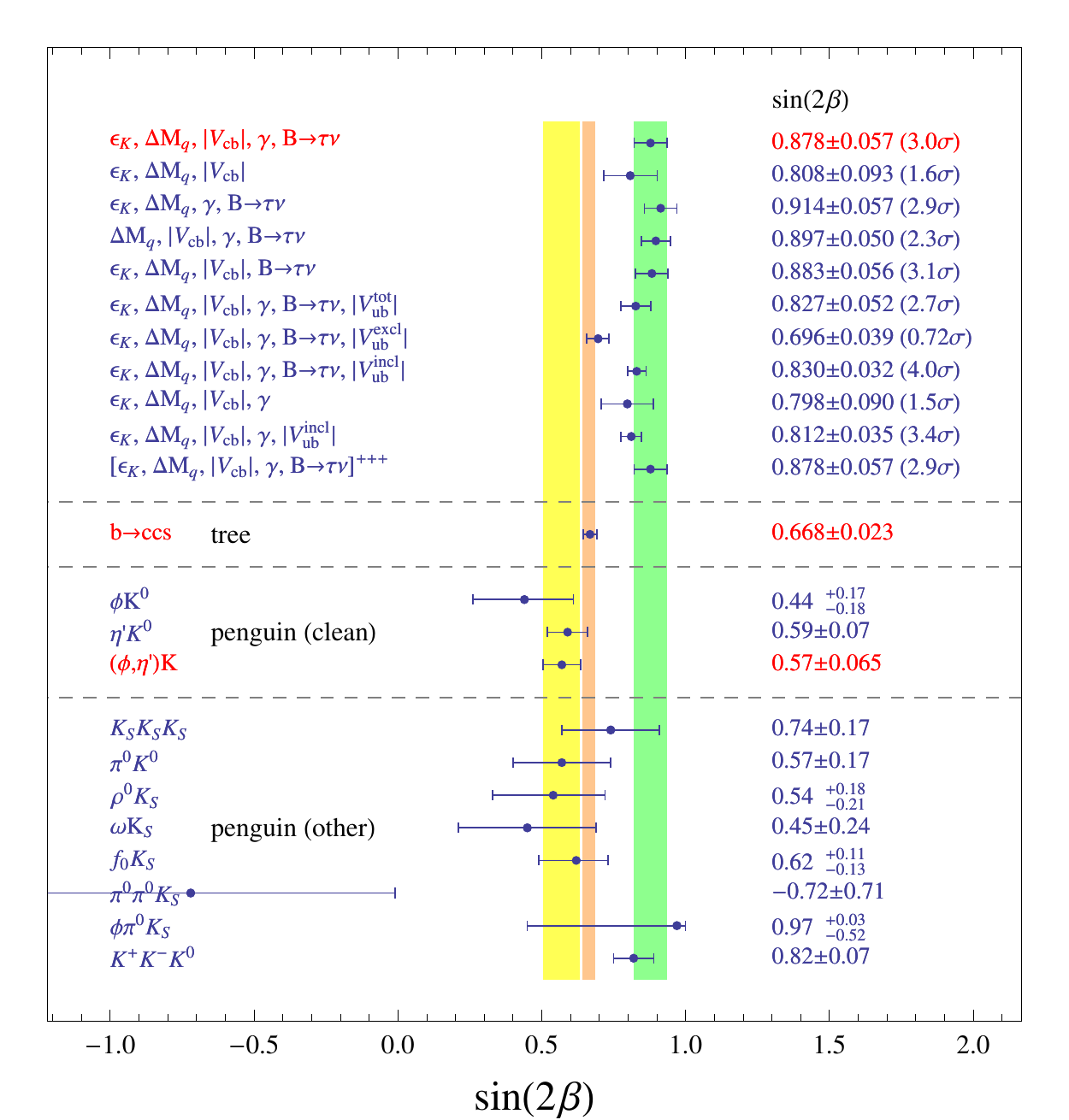}
\includegraphics[width=0.65 \linewidth]{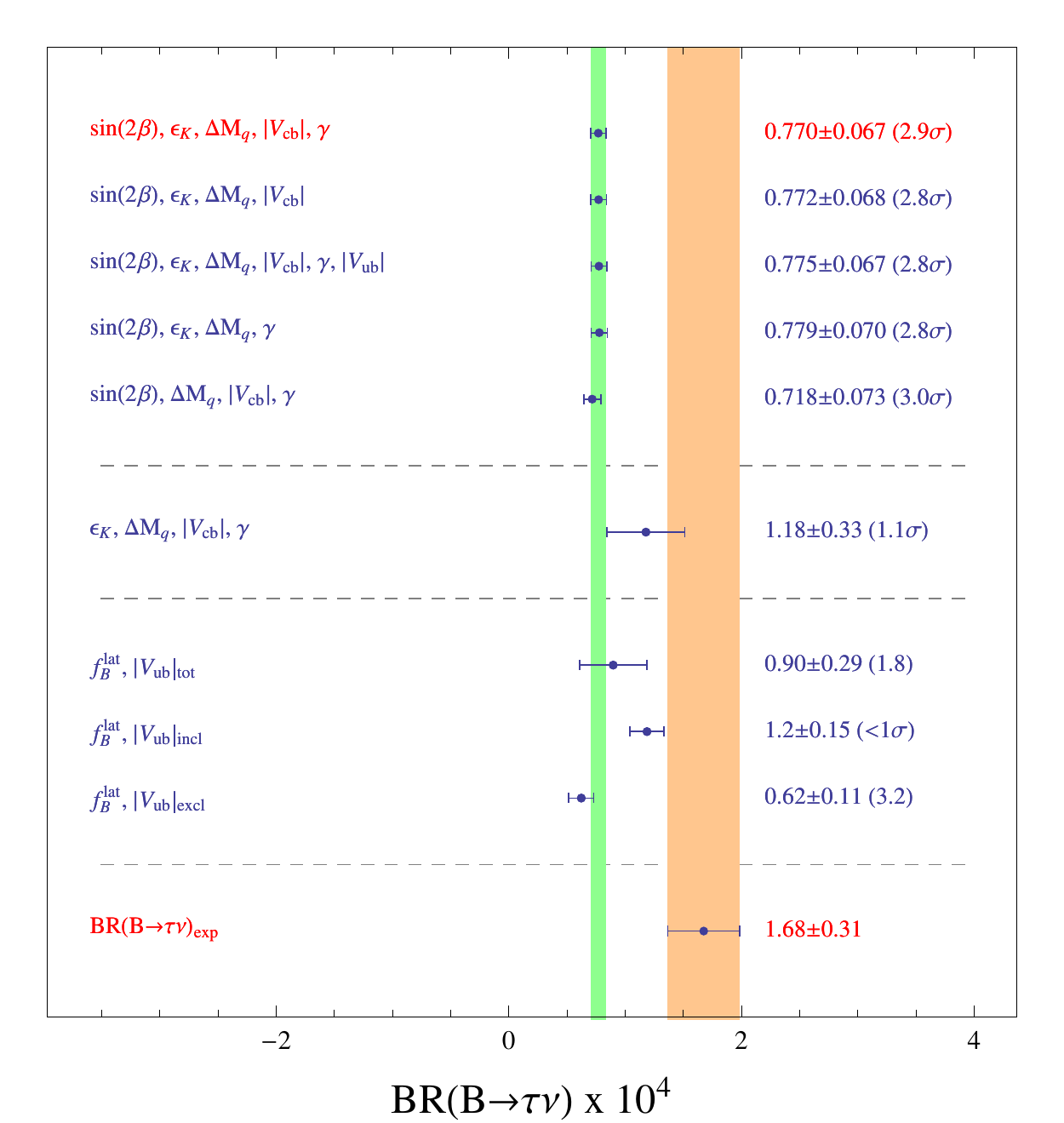}
\caption{{\bf Upper panel:} Summary of $\sin(2\beta)$ determinations. The entry marked +++ (tenth from the top) corresponds to adding an hadronic uncertainty $\delta \Delta S_{\psi K} = 0.021$ to the relation between $\sin (2\beta)$ and $S_{\psi K}$. {\bf Lower panel:} Summary of  ${\rm BR} (B\to\tau\nu)$ determinations. \hfill\label{fig:tabsin2beta}}
\end{center}
\end{figure}

\noindent
We collect the fit results obtained with different input selections and show the corresponding predictions for $\sin (2\beta)$ and ${\rm BR} (B\to \tau \nu)$ in Fig.~\ref{fig:tabsin2beta}. These tables are useful to get a clear picture of the stability of the observed tension against a variation of the inputs used.  The main lessons to be learned from the fits described in the previous sections are:
\begin{itemize}
\item The CKM description of flavor and $CP$ violation displays a tension at the 3$\sigma$ level, where most of the tension is driven by $B\to\tau \nu$ and $S_{\psi K}$.
\item The values of $|V_{ub}|$ obtained from inclusive and exclusive semileptonic $b\to u \ell \nu \; (\ell = e,\mu)$ differ by $3.3\sigma$.  This disagreement is cause for serious concern, and may indicate the presence of underestimated theoretical uncertainties.  Although the use of $|V_{ub}|$ in the UT fit is problematic, the tension persists even when the constraint from $|V_{ub}|$ is omitted.  A possibly related problem is that the Standard-Model prediction for ${\rm BR} (B\to \tau \nu)$ from $f_B$ and $|V_{ub}|_{\rm excl}$ (we obtain $(0.62 \pm 0.11) \times 10^{-4}$) is $3.2\sigma$ below the direct experimental measurement (see Table~\ref{inputs}). These two $3\sigma$ tensions in the $B\to \tau\nu$, $f_B$,  $|V_{ub}|$ system could be quite naturally resolved by a shift in the extraction of $|V_{ub}|$, either due to new physics or to improved theoretical understanding of inclusive and exclusive semileptonic $b\to u \ell\nu$ decays. 
\item  Once $|V_{ub}|$ is removed from the fit, the tension is driven by the disagreement between $B\to \tau\nu$ and $S_{\psi K}$, and can be alleviated by omitting either constraint.  In fact, the tension in the $S_{\psi K}$ prediction is reduced from $3.0\sigma$ to $1.5\sigma$ with the removal of $B\to \tau\nu$ (see the first and ninth entries in the upper table in Fig.~\ref{fig:tabsin2beta}).  Similarly, the tension in the ${\rm BR} (B\to\tau\nu)$ prediction is reduced from $2.9\sigma$ to $1.1\sigma$ with the removal of $S_{\psi K}$ (see the first and sixth entries in the lower table in Fig.~\ref{fig:tabsin2beta}).  Given the significant role played by BR($B\to \tau\nu$) in causing the tension in the UT fit, the possibilities of statistical fluctuations and/or of underestimated systematic uncertainties in the theoretical and experimental inputs to the constraint from $B\to\tau\nu$ should be given serious consideration.
\item    If we interpret the $3\sigma$ tension in the UT fit in terms of new physics, the fit prefers new contributions to $B\to\tau\nu$ and/or to $B_d$ mixing. Scenarios with new physics in kaon mixing are clearly disfavored (see the $p$-values in Eqs.~(\ref{cepsilon}-\ref{rh})).  In terms of a new-physics model whose interactions mimic closely the SM (see Eq.~(\ref{np})), this tension points to a few hundred GeV mass scale. Even allowing for a generous model dependence in the couplings, it seems that such new particles, if the tension in the fit stands confirmed, cannot escape detection in direct production experiments.
\end{itemize} 
\section*{Acknowledgements}
We would like to thank Christian Hoelbling and Urs Heller for their comments on this manuscript. We are especially grateful to Christine Davies for pointing out a missing renormalization factor in our interpretation of the $B^0_{(d,s)}$-mixing matrix elements.

\end{document}